\documentclass[aps,prd,twocolumn,showpacs,showkeys,nofootinbib]{revtex4-1}
\usepackage{graphicx}% Include figure files 
\usepackage{dcolumn}% Align table columns on decimal point 
\usepackage{amssymb} 
 
\usepackage{amsfonts} 
\usepackage{amssymb} 
\usepackage{amsmath} 
\usepackage{color}
 
\newcommand{\ar}{\arrowvert} 
\newcommand{\ra}{\rangle}

\newcommand{\be}{\begin{equation}} 
\newcommand{\ee}{\end{equation}} 
\newcommand{\ba}{\begin{eqnarray}} 
\newcommand{\ea}{\end{eqnarray}}

\begin{document} 
\title{Using QCD Counting rules to Identify the Production of  Gluonium}
\author{Stanley J. Brodsky $ ^1$ and Felipe J. Llanes-Estrada$ ^2$}
\affiliation{ 
$ ^1$ SLAC National Accelerator Laboratory, Stanford University, Stanford, CA 94025\\
$ ^2$Dept. F\'{\i}sica Teorica, Univ. Complutense, Madrid 28040, Spain
} 
 
\date{\today} 

%%%%%%%%%%%%%%%%%%%%%%%%%%%%%%%%%%%%%%%%%%%%%%%%%%%%%%%%%%%%%%%%%%%%%%%%%%%%%%%%%%%%%%%%%%%%%%%% 
\begin{abstract}  
%%%%%%%%%%%%%%%%%%%%%%%%%%%%%%%%%%%%%%%%%%%%%%%%%%%%%%%%%%%%%%%%%%%%%%%%%%%%%%%%%%%%%%%%%%%%%%%%
The empirical identification of bound states of gluons has remained a central goal of hadron spectroscopy. 
We suggest an experimentally challenging, but model--independent way to assess
which zero charge, isospin-zero mesons have a large gluonium light-front wavefunction component in  the quark and gluon Fock space of QCD.  
Our method exploits QCD counting rules which relate the power-law fall-off of production amplitudes at high momentum transfer to the meson's twist (dimension minus spin of its minimum interpolating operators). 
Scalar $0^+$ glueballs composed of two valence gluons with zero internal orbital angular momentum have twist $\tau=2$.
In contrast,  quark-antiquark $|q \bar q \rangle $ scalar mesons have twist $\tau \ge 3$  since they have nonzero orbital angular momentum, and multi-quark states
such as $|q  q \bar q \bar q \rangle$ tetraquarks yield twist $\tau \ge 4$.  Thus, the production cross section for both $|q\bar{q}\rangle$ and $|qq\bar{q}\bar{q}\rangle$ mesons will be suppressed by at least one power of momentum transfer relative to glueball production.   
For example, in single inclusive particle hadroproduction $ A B \to C X$, the cross section for glueball production at high transverse momentum $p_T$ and fixed $x_T = 2 {p_T\over \sqrt s} $ will dominate higher twist mesons by at least two powers of $p_T$. \\
Similarly, in exclusive production processes at large CM energy and fixed CM angle, the glueball rate dominates by a power of $s$: we illustrate the method with a simple reaction, $e^-e^+ \to \phi f_0$ where the $f_0$ can be tested to be a glueball versus another type of scalar meson.
\end{abstract} %end of abstract
%\pacs{
%} % end of PACS codes 
\maketitle

%%%%%%%%%%%%%%%%%%%%%%%%%%%%%%%%%%%%%%%%%%%%% 
\section{Introduction} \label{sec:intro}
%%%%%%%%%%%%%%%%%%%%%%%%%%%%%%%%%%%%%%%%%%%%% 

Since Quantum Chromodynamics (QCD) is a nonAbelian Yang-Mills theory, its force-carrying gluons interact nonlinearly,  
and thus it can in principle create gluonium bound states $|gg\rangle$ and $|ggg\rangle$ without quark constituents in its valence Fock state.
Such quark-less gluonium states, (often referred to as ``glueballs")  have been intensively studied by theorists; the consensus of the past two decades from lattice gauge theory~\cite{Albanese:1987pi,Morningstar:1999rf}, other approaches which model QCD, e.g., \cite{Sanchis-Alepuz:2015hma,Kuti:1998rh,Buisseret:2009yv,Szczepaniak:1995cw,Rinaldi:2018yhf}
as well as Regge theory~\cite{Kaidalov:1999yd,LlanesEstrada:2000jw},
is that the lightest $|gg\rangle$ glueball is a scalar $J^{PC}=0^{++}$ state in the $1.5-1.8$  GeV mass range, accompanied by a tensor $2^{++}$ state above 2 GeV (associated with the pomeron Regge trajectory).    Scalar $0^+$ glueballs composed of two valence gluon interpolating fields and zero internal orbital angular momentum have twist $\tau$ equal to $2,$   where twist is defined as the dimension minus spin of its minimum interpolating operators.  

QCD also predicts scalar mesons which are $|q \bar q\rangle $ P-wave bound states with internal  orbital angular momentum $L=1$ and twist $\tau\ge 3$,  as well as $0^{++}$ ``tetraquark" states $|q  q \bar q \bar q\rangle$ with twist $\tau \ge 4$.  The scalar $ |gg\rangle$ glueballs differ from these quark bound states by not having charge nor isospin, and by their twist $\tau =2$.   

The superconformal algebra approach~\cite{Dosch:2015nwa} to hadron physics predicts a unified spectroscopy of $|q\bar{q}\rangle$ mesons, quark-diquark baryons, and diquark/antidiquark bound $ |[qq] [\bar q \bar q ]\rangle$ tetraquarks as  members of the same 4-plet representation with a universal Regge slope.
However,  the application of superconformal algebra to QCD does not predict gluonic bound states: 
the very strong gluon-gluon self-interactions evidently lead to color-confining forces in the soft QCD domain, but not to the constituent gluon degrees of freedom underlying gluonium bound states. The experimental search for quark--less hadrons is thus a topic of central interest for QCD. 

Our method  for identifying gluonium states exploits QCD counting rules which relate the power--law fall--off of production amplitudes at high momentum transfer to the hadronic twist.   Similar counting rules for establishing tetraquarks and the composition of other exotic states have been presented~\cite{Brodsky:2015wza,Brodsky:2016uln}.

Experiments have identified a rich crop of scalar $f_0$ mesons in the 1-2 GeV energy interval~\cite{Patrignani:2016xqp}, at 980, 1370, 1500, 1710 and 2020 MeV. The BES $f_0(1810)$ candidate~\cite{Ablikim:2012ft,MartinezTorres:2012du} in $J/\psi\to \omega\phi$ could be the same as the $f_0(1710)$, distorted by phase space, so this leaves five scalar mesons in the region of interest, with two competing candidates often claimed to be glueballs, the 1500 and 1710.
However, the discussion of which of these two most closely resembles the theorized glueball is far from closed~\cite{Amsler:1995td,Giacosa:2004ug,Janowski:2014ppa}, with preference perhaps for the $f_0(1710)$. 

Several groups~\cite{Rosenzweig:1981cu,Cheng:2006hu,McNeile:2000xx,Narison:1996fm}  have  addressed~ the configuration mixing of glueballs with other ordinary or exotic mesons.
It  is clearly necessary to have clear criteria which bear on the two topics of glueball identification and  mixing. The large-$N_C$ expansion around $N_C=3$ partly provides such a criteria~\cite{Cohen:2014vta}; it can be tested in lattice gauge theory, but it would be more satisfactory to use experimental data directly.

Our observation is that the scalar glueball (almost certainly the lightest one) 
can be directly identified by experiment, albeit in a challenging kinematic regime, via  QCD counting rules.   We will briefly recall the basics of counting rules below and show how the identification can be carried out in exclusive reactions such as  $e^-e^+\to \phi f_0$ and other large transverse--momentum processes~\cite{Sivers:1975dg}.

%%%%%%%%%%%%%%%%%%%%%%%%%%%%%%%%%%%%%%%%%%%%%%%%%%%%%%%%%%%%%%%%%%%%%%%%%%%%%
\section{Counting rules and scalar glueball production} \label{BF}
%%%%%%%%%%%%%%%%%%%%%%%%%%%%%%%%%%%%%%%%%%%%%%%%%%%%%%%%%%%%%%%%%%%%%%%%%%%%%

An essential observation for a renormalizable theory is that, when all scales become large in an exclusive scattering process such as $AB\to CD$, at fixed CM angle and large Mandelstam-$s$), the differential cross section scales~\cite{Brodsky:1973kr,Matveev:1973ra} as a power-law in $s$, namely 
\begin{equation} \label{differentialcounting}
\frac{d\sigma(AB\to CD)}{dt}  = {f(\theta_{CM})\over {s^{n_i+n_f-2}}} .
\end{equation}
Here, $n_i$ and $n_f$ are the total minimum number of fundamental (pointlike) particles in the initial and final states (equivalently, the minimum number of fundamental fields necessary to interpolate between the vacuum and the initial and final scattering states).    Thus, for the reaction $e^-e^+\to \pi^+\pi^-$ one counts $n_i=2$ (two leptons in the initial state) and $n_f=4$ (each pion can be produced, at a minimum, from a quark-antiquark pair). 
This yields $\frac{d\sigma}{dt} \propto 1/s^4$ which corresponds, after accounting for kinematic factors, to a pion form factor $F_\pi(s)\propto 1/s$, a prediction under intense study~\cite{Chang:2013nia}.   The light-front Drell-Yan-West formulae for electromagnetic and gravitational  form factors are identical to the Polchinski-Strassler~\cite{Polchinski:2001tt} formula for form factors in Anti-de Sitter space (AdS$_5$).  This identification ({\it light-front holography}) also provides a nonperturbative derivation of the scaling laws for form factors at large momentum transfer.  
Numerous other predictions~\cite{Brodsky:1983vf}, including helicity selection rules~\cite{Brodsky:1981kj}, have been put forward for exclusive processes.
The power--law predictions acquire logarithmic corrections as predicted in pQCD using the ERBL evolution equation~\cite{Lepage:1980fj,Efremov:1979qk}.

The less-used extension that we need is the inclusion of orbital angular momentum~\cite{Amati:1968kr,Ciafaloni:1968ec,Brodsky:1974vy}. 
Just like the nonrelativistic wavefunction of a bound state is damped at short distances by a centrifugal factor $r^L$, the light-front wavefunctions and the Bethe-Salpeter wavefunctions contain also such suppression. In the front form, the corresponding boost-invariant ``radial" variable is $\zeta$ where $\zeta^2 = b^2_\perp x(1-x)$, and $J^z = L^z + S^z$ is conserved at every vertex~\cite{Chiu:2017ycx} 
This means that amplitudes involving a hadron with $L$ units of internal angular momentum are suppressed by $\left({\sqrt s}\right)^{-L}$~\cite{Brodsky:1981kj}, and the cross sections by $s^{-L}$.
As a consequence, the counting rules reflect the hadron twist $\tau$ and the cross section become
\begin{equation} \label{counting}
\frac{d\sigma}{dt}  = {f(\theta_{CM})\over {s^{n_i+n_f+L -2}}} .
\end{equation}
where $L$ sums all the internal orbital angular momenta.

Let us then apply the counting  rules to the identification of a glueball among the $f_0$ states.
The minimum Fock state that can appear in a glueball with $J^{PC}=0^{++}$ is
$\ar \vec{g}\cdot \vec{g} \rangle$ with the gluon spins antialigned and no orbital angular momentum. Thus, $n_f+L=2$: see table~\ref{tab:suppression}. 

\setlength{\tabcolsep}{12pt}
\renewcommand{\arraystretch}{1.5}
\begin{table}
\caption{Power of $s$ in the QCD counting rules suppressing the production of other wavefunctions \emph{relative to the glueball} in large momentum transfer reactions involving an $f_0$ meson.\label{tab:suppression}}
\begin{tabular}{|c|cccc|} \hline
Wavefunction& $gg$  & $q\bar{q}\arrowvert_{L=1}$ & $q\bar{q}g$ & $q\bar{q} q\bar{q}$ \\
$n_f+L$     &  2    &  3         & 3         & 4 \\               
Suppression &  1    & $s^{-1}$   & $s^{-1}$  & $s^{-2}$ \\
\hline 
\end{tabular}
\end{table}

The table also shows various other configurations that can also appear in a scalar, isoscalar meson which are power-law  suppressed in exclusive, large momentum transfer reactions. Adding extra fields  further suppresses the production cross section. In the next section~\ref{sec:exreaction} we illustrate the counting rules for a simple exclusive  $e^+ e^-$ annihilation process.

%%%%%%%%%%%%%%%%%%%%%%%%%%%%%%%%%%%%%%%%%%%%%%%%%%%%%%%%%%%%%%%%%%%%%%%%%%%%%
\section{Example reaction: $e^-e^+\to \phi f_0$\label{sec:exreaction}}
%%%%%%%%%%%%%%%%%%%%%%%%%%%%%%%%%%%%%%%%%%%%%%%%%%%%%%%%%%%%%%%%%%%%%%%%%%%%%

\begin{figure}
\includegraphics[width=0.45\columnwidth]{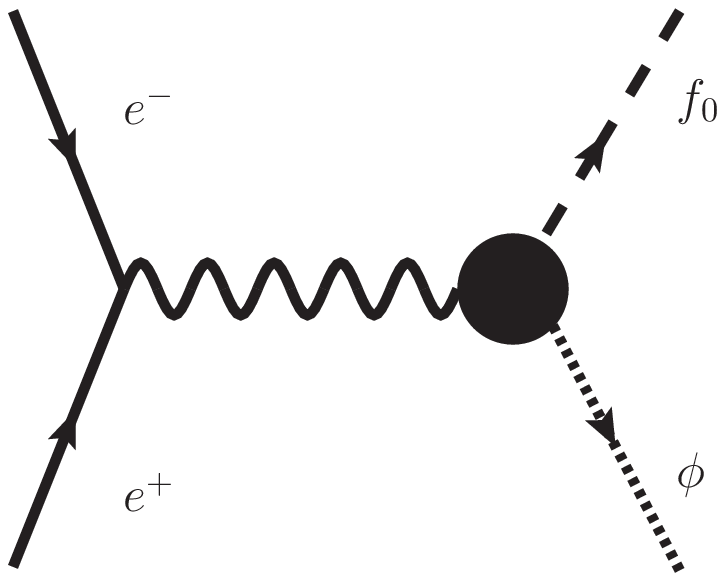}\hspace{3mm}
\includegraphics[width=0.5\columnwidth]{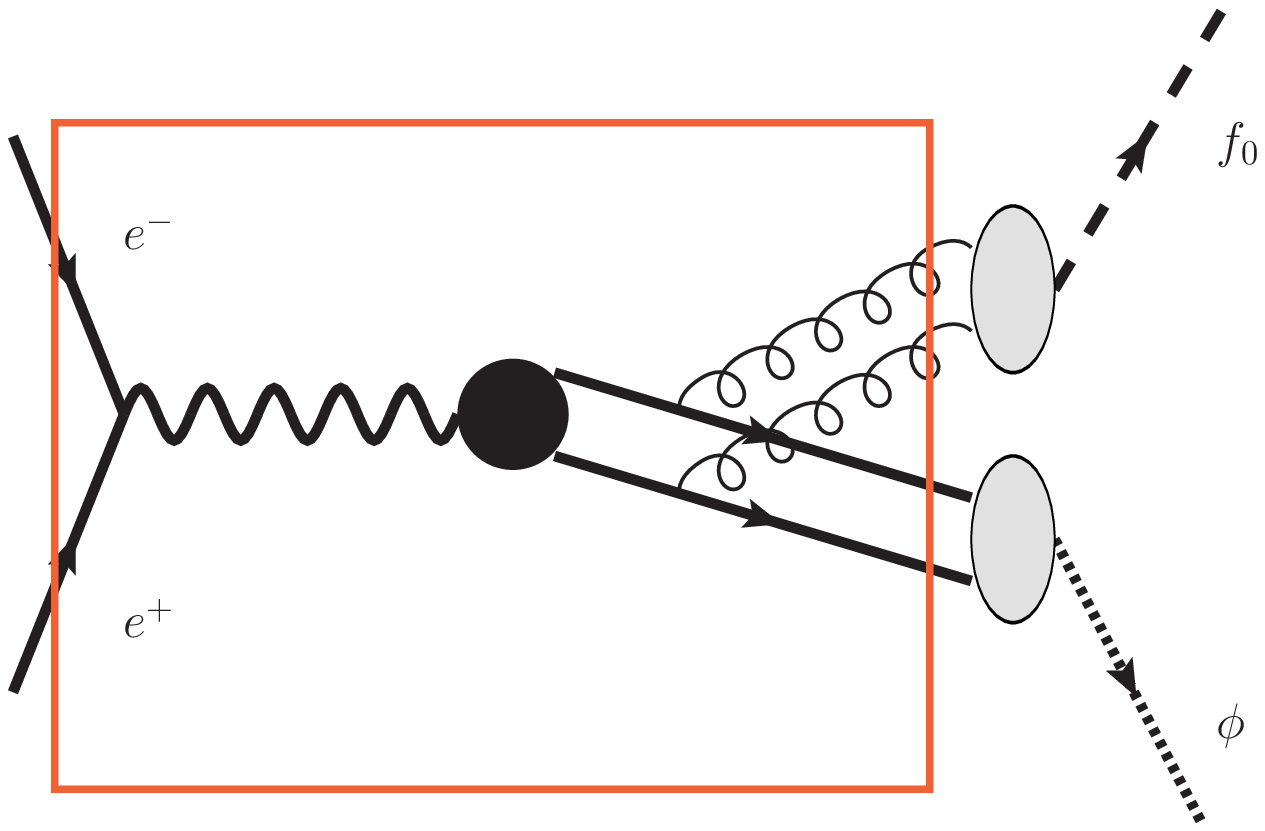}
\caption{\label{fig:Feynmanhadrons} A process which can distinguish the scalar glueball among the $f_0$s: $e^-e^+\to\phi f_0$ at the hadron level and at the quark level. In the right diagram, the counting rules correspond to the number of underlying fundamental fields (lines intersecting the box, red online) in the initial and final states.}
\end{figure}

Exclusive reactions involving large transverse momentum transfer or $t$ are challenging because their cross sections fall as a power law against a background of total hadronic cross sections which are logarithmically growing. However, modern detectors in high luminosity machines, such as Belle-II, can provide good identification against large backgrounds.  Among many similar exploitable reactions, we exemplify the advantageous process $e^-e^+\to \phi f_0$ (see Feynman diagram in figure \ref{fig:Feynmanhadrons}).

Indeed, the $\phi$ recoiling against the scalar meson has a basic $s\bar{s}$ quark composition (ideal mixing) in an $L=0$ state that makes the application of the counting rules straightforward. 
The two mesons eventually decay with a sizeable branching fraction to $K^-K^+$ and $\pi^+\pi^-$, four charged tracks which are easily identifiable in Belle's barrel detector. Since this is an exclusive process,  no other particle can appear in the same event.
The counting rule of Eq.~(\ref{counting}) can then be applied (see right plot of figure~\ref{fig:Feynmanhadrons}): $n_i=2$ for the leptons, $n_f=4$ for a quark-antiquark pair and two gluons, and $L=0$, yielding $\frac{d\sigma}{dt}  = f(\theta) \frac{1}{s^4}$.

Counting all events in the barrel detector amounts to integrating over a fixed solid angle ($t$ not suppressed respect to $s$), and all scales are large. Then,
%\begin{equation} 
$\sigma\arrowvert_{\rm barrel} = 4\ar {\bf p}_\phi\ar \ar {\bf p}_{f_0}\ar 
\times \int_{0}^{\cos\theta_{\rm min}}  \! \! \! d\cos\theta \ \ \frac{d\sigma}{dt}
$
%\end{equation}
adds one power of $s$, resulting in the asymptotic behaviors (up to logarithms),

\begin{eqnarray}\label{glueballscaling}
\sigma \left(f_0=\ar \bf{gg} \ra +\dots \right) & \sim & \frac{\rm constant}{\bf s^3} \\ \nonumber
  & \phantom{\sim} & \\ \nonumber
\sigma \left(f_0=\ar {\bf{q\bar{q}}}\ra_{L=1} +\dots \right)       & \sim& \frac{\rm constant}{\bf s^4}   \\ \nonumber
  & \phantom{\sim} & \\ \nonumber
\sigma \left(f_0=\ar {\bf{q\bar{q}q\bar{q}}}\ra_{s-{\rm wave}} +\dots \right)       & \sim& \frac{\rm constant}{\bf s^5}   
\end{eqnarray}

Belle-II could then measure this reaction, {\it e.g.} at 9 and 11 GeV (off--resonance to avoid complications from $\Upsilon(b\bar{b})$ structure). The ratio of the reaction cross sections at the two energies would fall by a factor, depending on the quark and gluon valence composition of the $f_0$, given by
$ \frac{\sigma(9{\rm GeV})}{\sigma(11{\rm GeV})} \simeq 3.4 \ (gg)\ ; \ 5 \ (q\bar{q})_{L=1}\ ; \ 
7.5\ (qq\bar{q}\bar{q})$, etc.
Thus, a measurement of this cross--section ratio to 20\% accuracy can provide a meaningful test. 
Because the quark-gluon composition of the various $f_0$s are different, the spectrum is distorted by those factors as the collider energy increases, as we show in the next paragraph~\ref{sec:counts}.
  
The isoscalar gluonium production can be confirmed by verifying that no charged p-wave state with twist $\tau = 2$ appears at the same mass in channels such as $e^+ e^- \to \rho^\pm  a^\mp$.

%%%%%%%%%%%%%%%%%%%%%%%%%%%%%%%%%%%%%%%%%%%%%%%%%%%%%%%%%%%%%%%%%%%%%%%%%%%%%%%%%%%%%%%%%%%%
\section{Event number estimate}\label{sec:counts}
%%%%%%%%%%%%%%%%%%%%%%%%%%%%%%%%%%%%%%%%%%%%%%%%%%%%%%%%%%%%%%%%%%%%%%%%%%%%%%%%%%%%%%%%%%%%

\begin{figure}
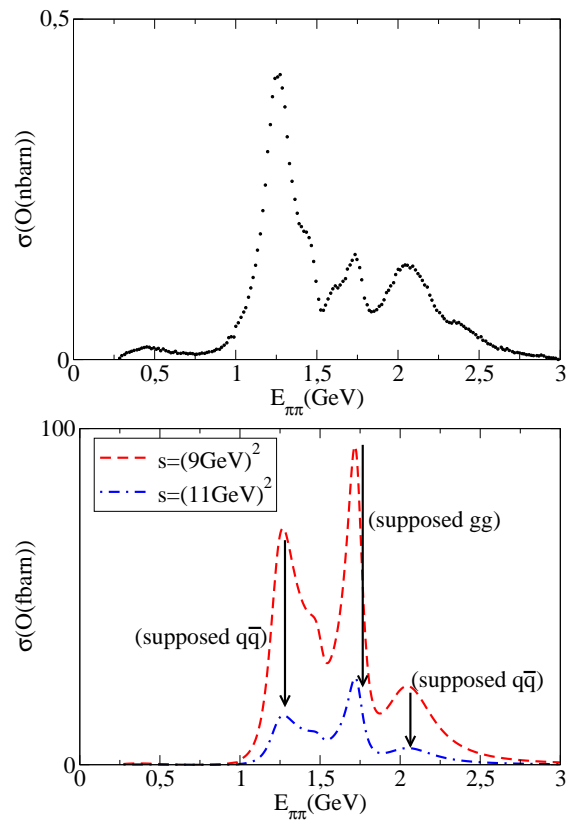

\includegraphics*[width=0.85\columnwidth]{FIGS.DIR/Jpsitopipiphoton.eps}
\includegraphics*[width=0.85\columnwidth]{FIGS.DIR/pipispectrum.eps}
\caption{\label{fig:spectrumdistort} 
Top: Experimental $\pi\pi$ spectrum at the 3 GeV scale obtained~\cite{Bennett:2014fgt} in $J/\psi \gamma \pi\pi$.
Bottom: an example of the $\pi\pi$ spectrum resulting from 
$e^-e^+\to \phi f_J$ at 9 and 11 GeV under the assumption that $f_0(1710)$ is the glueball
(the state that drops the least in the measurement of the rate, if consistent with Eq.~(\ref{glueballscaling}), would fit this assignment).}
\end{figure}

The well known $C=+1$ $\pi\pi$ spectrum from radiative $J/\psi$ decays~\cite{Bennett:2014fgt} is shown in the top plot of figure~\ref{fig:spectrumdistort}. The typical scale here is thus at the charmonium's 3.1 GeV~\footnote{We take this energy to be the watershed between strongly interacting hadron physics (with the same spectrum in all reactions, as per Watson's final-state theorem) and the hard regime germane to the QCD counting rules. Choosing a higher energy increases the predicted number of events (our result should then be a lower bound), because in the hadronic regime the cross--section falls less steeply, as argued next in section~\ref{sec:hadron}; but the spectrum at high energies is less distorted than shown in fig.~\ref{fig:spectrumdistort}.}. 

Having the line shape at hand, we need to normalize the spectrum at the same $\sqrt{s}$.
We profit from actual Belle and Babar measurements~\cite{Shen:2009mr}
of $e^-e^+\to \phi \pi\pi$ at the $f_0(980)$ mass, which fixes the total cross section at one point in the spectrum.
The cross section at low energy is dominated by the $Y(2175)$ and peaks around 0.6 nb, then falling to about 0.05 nb at 3 GeV, with no sign of significant resonances. We take this last number as our normalization of the $OY$ axis in fig.~\ref{fig:spectrumdistort}, and assume for the estimate that the counting rules apply for $E>3$ GeV. 

We can then use the power--laws of Eq.~(\ref{glueballscaling}) to estimate the cross--section under various scenarios. The lower panel of fig.~\ref{fig:spectrumdistort} assumes, for the sake of demonstration, that $f_0(1710)$ is mostly the glueball and the other visible $C=+1$ states, saliently the $f_2(1270)$, behave as a $q\bar{q}$ meson. 
With $\sigma(9{\rm GeV})$ reduced to $\sim70$ fbarn, Belle-II could produce some 70000 $\phi$--recoiling $f_0(1710)$s with 1 ab$^{-1}$ of integrated luminosity (just 20\% of a year's data taking). About 20000 events are also obtainable at 11 GeV.

%%%%%%%%%%%%%%%%%%%%%%%%%%%%%%%%%%%%%%%%%%%%%%%%%%%%%%%%%%%%%%%%%%%%%%%%%%%%%%%%%%%%%%%%%%%%
\section{Reaction at the hadron level}\label{sec:hadron}
%%%%%%%%%%%%%%%%%%%%%%%%%%%%%%%%%%%%%%%%%%%%%%%%%%%%%%%%%%%%%%%%%%%%%%%%%%%%%%%%%%%%%%%%%%%%

The $e^+ e^-$ annihilation cross section at high $s$ (very virtual photon) is
$ \left.\frac{d\sigma}{d t}\right|_{\rm CM} = \frac{1}{16\pi s^2} 
\overline{\arrowvert {\mathcal{M}} \arrowvert^2}$, where 
the squared, spin-averaged Feynman amplitude is
$ \overline{\arrowvert {\mathcal{M}} \arrowvert^2} = \frac{1}{4} \frac{e^4}{s^2}
L^{\mu\nu}H_{\mu\nu} $
in terms of the $e^-e^+$ lepton tensor $L^{\mu\nu}=k_-^\nu k_+^\mu + k_-^\mu k_+^\nu - \frac{s}{2}\eta^{\mu\nu}$ and the hadron tensor $H$. There is only one independent scalar variable that $H$ can depend on -- the virtuality of the photon $q^2=s=Q^2$, as both final state particles are on-shell ($p_i^2=M_i^2$, $i=\phi,\ f_0$) and 4-momentum is conserved, $q^2=(p_\phi+p_{f_0})^2$.

As for the fields, we have three at the vertex (blob in the left plot of fig.~\ref{fig:Feynmanhadrons}), a scalar one $f_0$ and the two vector ones that satisfy, $\partial_\mu A^\mu = 0 = \partial_\nu \phi^\nu$, so that the effective Lagrangian must contain the indices crossed, {\it i.e.} $\partial_\mu \phi_\nu$ etc. 
Because $\mathcal{L}$ is a scalar and because of conservation of parity, there can only be terms with an even number of derivatives and no Levi-Civita antisymmetric symbols. The Klein-Gordon operator acting on the final state particles can be substituted by a constant, as $\Box f_0 = m^2_{f_0} f_0$ (and similarly for $\phi$). Leibniz's rule for the derivative of a product and the neglection of total derivatives further reduce the interaction vertex to two components,
\begin{equation}
V_\mu = F_1(q^2) \epsilon(\phi)_\mu + F_2(q^2) (q\cdot \epsilon(\phi))p(\phi)^\mu
\end{equation}
in terms of the polarization $\epsilon(\phi)$ and momentum $p(\phi)$ of the final state $\phi$ meson and two unknown form factors $F_1$ and $F_2$ that control its asymptotic behavior when $s\sim t\sim u \to \infty$. At high momentum transfer, the squared, unpolarized  Feynman amplitude becomes
\begin{equation}
\overline{\arrowvert {\mathcal M}\arrowvert^2} = \frac{e^4}{4s^2}\left[
\frac{tu}{2m_\phi^2}\left( \arrowvert F_1\arrowvert^2 +\frac{s^2}{4} \arrowvert F_2\arrowvert^2
+\frac{s}{4}(F_1^*F_2+F_2^*F_1) \right)\right]\ ;
\end{equation}
(diagonalization by rotating the $F$s is unnecessary now).

In the perspective of effective field theory, hadrons at low momentum transfer act as if pointlike and 
can be treated with and effective Lagrangian, providing a baseline against which to judge the quark-gluon based results.

Adopting the Vector Meson Dominance model, in which the photon fluctuates into a vector meson (mostly a $\phi$--like state such as $\phi(1680)$ or $Y(2175)$) and constructing an interaction Lagrangian along the lines of~\cite{Black:2006mn}, 
\begin{equation}
{\mathcal L_{\phi'\phi f_0}} = \frac{\beta}{2}f_0(\phi'_{\nu,\mu}-\phi'_{\mu,\nu})(\phi^{\mu,\nu}-\phi^{\nu,\mu}) + \frac{e}{3} \tilde{g} f_\pi^2 A^\mu \phi'_\mu\ , 
\end{equation}
the form factors can be seen to behave as 
\begin{equation}
F_1(s) \to {\rm constant}=\frac{e}{3}\tilde{g}f_\pi^2\beta\ , \ F_2(s) \to - \frac{F_1(s)}{2s} \ .
\end{equation}
This is the pointlike--hadron limit, distinct from the QCD prediction for 
the timelike form factor of an s--wave $\ar q\bar{q}\rangle$--meson,  up to logarithms $F_1\sim 1/s$,
that also applies to glueball--$\phi$ production.
Thus, while the least  drop in $\sigma$ that QCD may support for large $s$ is $1/s^3$ as in Eq.~(\ref{glueballscaling}), if the $\phi$ and $f_0$ are taken as pointlike hadrons, the cross--section falls much more slowly at high $s$ as given by
\begin{equation}
\sigma_{\rm hadronic}(e^-e^+\to \phi f_0) \propto \frac{1}{s}\ .
\end{equation}
This result~\footnote{Actually, the counting rules for two structureless mesons predict, with $n_i=2=n_f$, that  $d\sigma/dt \sim 1/s^2$ or, integrating over the angular span of the barrel, exactly this behavior $\sigma\sim 1/s$. The rules encode naive dimensional analysis that the model respects.} can be used in the experiment as the null hypothesis (no access to the internal structure of the hadron) and shows that the number of events for the $f_0(1710)$ estimated in section~\ref{sec:counts} is a lower bound.

%%%%%%%%%%%%%%%%%%%%%%%%%%%%%%%%%%%%%%%%%%%%%%%%%%%%%%%%%%%%%%%%%%%%%%%%%%%%%%%%%%%%%%%%%%%%%%%
\section{Effect of the renormalization of the wavefunction}
%%%%%%%%%%%%%%%%%%%%%%%%%%%%%%%%%%%%%%%%%%%%%%%%%%%%%%%%%%%%%%%%%%%%%%%%%%%%%%%%%%%%%%%%%%%%%%%

The experiment we propose can reveal the states with large glueball wavefunction components at a hard scale of $\Lambda_1=9$ GeV. But one question that may arise is the effect on the Fock expansion of a given state upon changing the scale down to a typical hadronic $\Lambda_2=3$ GeV. 

Wavefunctions are not renormalization-scale invariant objects, but we recall~\cite{Burkardt:2002uc}
that the entire scale dependence of the light cone wavefunction is factorized into field renormalization constants: for example, the $q\bar{q}$ component is 
$\varphi^\Lambda(x,k_\perp) = Z_2(\Lambda) \tilde{\varphi}(x,k_\perp)$ in terms of the light--front variables, and where $\tilde{\varphi}$ is scale--independent.  The glueball wavefunction of interest here is 
\begin{equation}
\varphi^\Lambda_{gg}(x,k_\perp) = Z_A(\Lambda) \tilde{\varphi}_{gg}(x,k_\perp)\ ;
\end{equation} 
for a state with $n_q$ quarks and antiquarks and $n_g$ gluons, the constant is
$Z_2^{n_q/2}Z_A^{n_g/2}$.
This means that, for the glueball component,
$ \frac{\varphi^{\Lambda_2}_{gg}}{\varphi^{\Lambda_1}_{gg}} = \frac{Z_A(\Lambda_2)}{Z_A(\Lambda_1)} $.
But the quark and gluon field renormalization constants are known~\cite{Lappi:2016oup} to vary with the scale as (to one loop and ignoring $\log(\alpha_s)$) 
\begin{eqnarray}
Z_2 \simeq 1 +  {\rm div.}
- \frac{3}{2} \frac{g^2C_F}{8\pi^2}\log \left(\frac{\mu^2}{\Lambda^2}\right) +\dots \\
%Z_A \simeq 1 +  {\rm div.} +\dots + \\ \nonumber  \frac{g^2}{8\pi^2}\left(
%\left(\frac{11C_A}{6}-\frac{2T_FN_f}{3}\right) \log\left(\frac{\mu^2}{\Lambda^2}\right) 
%-C_A \frac{5+6\pi^2}{36}
%\right)
\!\! Z_A\! \simeq\! 1\! +\!  {\rm div.}\! +\!  \frac{g^2}{8\pi^2}\!
\left(\!\!\frac{11C_A}{6}\!-\!\frac{2T_FN_f}{3}\!\!\right)\! \log\!\left(\!\frac{\mu^2}{\Lambda^2}\!\right)\!+\!\dots 
\end{eqnarray}
With $\alpha_s\simeq 0.19$ at the 9 GeV scale and choosing $\mu\equiv\Lambda_2=3$ GeV, 
the change in $Z_A$ is due to $0.116\log\left( \frac{3^2}{9^2}\right)\simeq -0.255$. In turn, the difference of quark renormalization constants is +0.133.
This means that the glueball wavefunction of a mixed state determined at 9 GeV can decrease by order 30\% by the time the hadron scale is reached; the pure $q\bar{q}$ wavefunction takes an increase of order 13\%. Nevertheless, the renormalization corrections are multiplicative: very small wavefunction components remain very small. 
In the lucky event that only one state contained most of the glueball in the hard-scale experiment, this state would still contain it at the lower scale.

%%%%%%%%%%%%%%%%%%%%%%%%%%%%%%%%%%%%%%%%%%%%%%%%%%%%%%%% 
\section{A comment on other exotic scalars below 2 GeV}
%%%%%%%%%%%%%%%%%%%%%%%%%%%%%%%%%%%%%%%%%%%%%%%%%%%%%%%%
From the point of view of the counting rules, table~\ref{tab:suppression} shows that 
$|q\overline{q}g\rangle$ hybrid mesons follow the same power--law as ordinary p--wave $|q\overline{q}\rangle$ mesons, so they cannot be distinguished; additionally light scalar hybrids are not generally expected below 2 GeV. Therefore, we will comment only briefly on tetraquark states.

Since the classic work of Jaffe~\cite{Jaffe:1977cv}, a nonet of light scalar mesons is expected.  Precision studies have been carried out for the $\sigma$ or $f_0(500)$~\cite{Pelaez:2015qba} and $\kappa$ or $K^*_0(700)$ mesons: they lead to the belief that these states 
complete such nonet together with the $f_0(980)$ and $a_0(980)$.

Moreover, the recent realization~\cite{Nielsen:2018uyn} that an approximate supersymmetry among the meson, baryon and tetraquark spectrum may be at work, naturally leads to the assignment of the octet formed by $\sigma$, $\kappa$, and $a_0(980)$, as the superpartner of Gell--Mann's baryon $N(940)$, $\Lambda(1110)$, $\Sigma(1190)$, $\Xi(1320)$ octet. Then, given the mass similarity and proximity to the $K\bar{K}$ threshold of $f_0(980)$ and $a_0(980)$, it is reasonable to think that $f_0(980)$ is of the same tetraquark--like nature. 
%%%%

If this assignment is correct, then both $f_0(500)$ and $f_0(980)$ should 
have a fast dropping cross--section between 9 and 11 GeV (a factor 7.5 as explained at the end of section~\ref{sec:exreaction}). In fact, if the counting rules apply from as low as $E=3$ GeV, the relative drop of a $|qq\bar{q}\bar{q}\rangle$ candidate is so large (a factor of (3 GeV/9 GeV)$^4$=1/81 with respect to the glueball production rate, 1/9 respect to the quark-antiquark state) that it would be erased from the spectrum.

Any subdominant $|q\bar{q}\rangle$ components of the (mostly tetraquark) light $f_0$ states~\cite{Pelaez:2015qba} would come to the front, so that $\sigma\sim \frac{a_4}{s^4}+ \frac{a_3}{s^3}$ with $a_3<a_4$ would eventually become dominated by $a_3$.  
A possible experimental outcome is that after a quick change of the spectral shape due to erasing the tetraquark components out of the light $f_0$s, eventually mixed states would be decreasingly produced in pace with the largely $q\bar{q}$ $f_2(1270)$. 

As for further tetraquark states in the 1-2 GeV energy interval, we should recall that tetraquarks generically form flavor multiplets. 
There is~\cite{Patrignani:2016xqp} a visible $a_0(1450)$ that can be assigned to the same $f_0(1370)$ (largely $|q\bar{q}\rangle$) multiplet including also either the $f_0(1500)$ or $f_0(1710)$ as the $|s\bar{s}\rangle$ partner, but one or a linear combination of these $f_0$s is an isoscalar supermultiplet, so it is more likely a glueball than a tetraquark. 
There possibly is an $a_0(1950)$ to match $f_0(2020)$, so there could be an excited multiplet, either of $|q\bar{q}\rangle$ or $|qq\bar{q}\bar{q}\rangle$. The counting rules can help discern the nature of all these states by following the behavior of their production cross--section with energy. 

%%%%%%%%%%%%%%%%%%%%%%%%%%%%%%%%%%%%%%%%%%%%%%%%%%%%%%%%%%%%%%%%%%
\section{Glueball production in $pp$ collisions}
%%%%%%%%%%%%%%%%%%%%%%%%%%%%%%%%%%%%%%%%%%%%%%%%%%%%%%%%%%%%%%%%%%

Our arguments can be extended to other high momentum transfer exclusive and semi-inclusive reactions. 
In this paragraph we  briefly address the counting rules for proton--proton collisions which could be carried out at RHIC or by the CMS or ALICE collaborations at the LHC.
In this case both protons scattered elastically (e.g,  to roman pots set at fixed angles along the beam pipe) as in $pp\to pp \phi f_0$. The meson subsystem is deposited in the central barrel with a sizeable transverse momentum (2-5 GeV for each meson) so that pomeron and other Regge exchanges subside. In the case where all angular intervals are fixed and the large momentum transfer scales are large  we can apply the counting rules.
The proton's elastic scattering brings a  decreasing form factor, proportional to $1/q^4$. The counting rules predict  $d \sigma/dt \sim s^{2-n_i-n_f} = s^{-14}$   for $f_0\sim gg$    and $s^{-15}$ for 
$|q\bar{q}\rangle$ ($s^{-16}$ in the tetraquark case), since the protons provide six particles in each of the initial and final states. 
Such a strong fall-off is not going to be easily distinguishable (a precision under 10\% is required in the measurement of the exponent at 5-10 GeV!).
Therefore, we additionally propose  doubly diffractive peripheral two-photon measurements with large $p_t$ (of several GeV for each of the two mesons in the barrel, where a double gap ensures that both protons are diffracted). Since the protons are not required to scatter elastically, there is no power--law suppression from their quark content. Large $p_t$ is required to ensure that the particles extracted from the proton are pointlike (typically photons) so that pomeron-reggeon exchanges play no role.
Under these conditions, the prediction is identical to the one in $e^-e^+$ and $\gamma \gamma$ annihilation, since the effective reaction is 
$\gamma\gamma \to \phi f_0$:  the initial state, in practice, is made of two pointlike particles.
At any fixed energy the cross section will be relatively small because of the electric--charge dependent extraction of the proton, the diffractive
requirement on the protons, and the large $p_t$ requirement on the mesons. 
But once this has been accounted for, the power--law suppression
of $d\sigma/dt$ is much less steep and more easily accessible.

These measurements beyond our proposed  reaction  in section~\ref{sec:exreaction}
do seem promising, and we intend to focus future studies estimating their feasibility.   

%%%%%%%%%%%%%%%%%%%%%%%%%%%%%%%%%%%%%%%%%%%% 
\section{Outlook}
%%%%%%%%%%%%%%%%%%%%%%%%%%%%%%%%%%%%%%%%%%%% 

\begin{figure}
\includegraphics*[width=0.9\columnwidth]{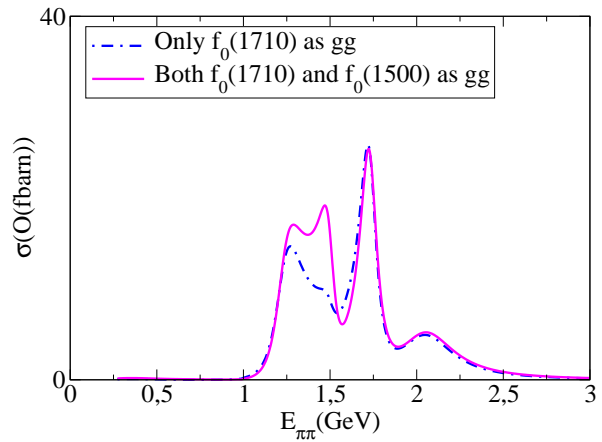}
\caption{\label{fig:mixing} A possible $\pi\pi$ line shape in the reaction
$e^-e^+\to \phi\pi\pi$ at 11 GeV. The broken line shows a simple model in which only the $f_0(1710)$ contains a significant glueball component, and thus $\sigma\sim s^{-3}$ from Eq.~(\ref{glueballscaling}). The solid line recalculation assumes that the $f_0(1500)$ also scales as a glueball (both states, strongly mixed, share the glueball): $f_0(1500)$ ends up dominating over the $f_2(1270)$ peak, assumed to be a $|q\overline{q}\rangle$.}
\end{figure}

In addition to the exclusive measurements, one can consider the one particle inclusive reaction $e^-e^+\to \phi + X$, analogous to the charmonium one~\cite{Abe:2007jna} that was used to discover the $X(3940)$. Then, the $\pi^+$ $\pi^-$ would not need to be reconstructed, as long as the $\phi$ be hard. This would increase the counting rate as the neutral decays of the recoiling $f_0$ would be included.
Multiparticle decays of the $f_0$ would not change the counting here, as the most likely quasi-collinear emission, not involving another highly virtual particle, does not cost an additional power of $s$, as recently emphasized~\cite{Brodsky:2017icd}.

If the mixing angle rotating the $|q\overline{q}\rangle$ and $|gg\rangle$ to the physical $f_0$ basis is large. Then, more than one state may follow the glueball counting rule of Eq.~(\ref{glueballscaling}). The situation is illustrated in figure~\ref{fig:mixing}. We see that the experimental line shape can be used to decide about this scenario.   For example, if the $f_0(1500)$ is taken to have a sizeable glueball fraction, it eventually becomes more prominent than the (likely $q\overline{q}$) $f_2(1270)$ that towers the spectrum at low-$s$, and with which it seems to interfere.

We emphasize that Belle-II can make an important contribution to hadron spectroscopy by identifying exotic hadronic states, including glueballs and tetraquarks~\cite{Drutskoy:2012gt,Kou:2018nap}.
If Belle-II collects significant off-resonance data at 9 and 11 GeV (or other sensibly chosen energies), it can make a fundamental test of the nature of the $f_0$ mesons and help with a longstanding puzzle, the identification of the glueball.
Moreover, any scalar meson $f_0$ which has an $O(1)$ mixing overlap with a glueball will have  $\sigma(e^+ e^- \to \phi f_0)$ scaling as $1/s^3$; thus, Belle can \emph{experimentally prove} the existence of a glueball even if it strongly mixed among several states, by just identifying a fraction of the spectrum with that specific scaling.

This procedure  can be extended to the tensor $2^{++}$ glueball which is expected to have a mass slightly above 2 GeV; such quantum numbers can also arise from a $p$--wave $|q\bar{q}\rangle$ wavefunction, which can be distinguished from the $|gg\rangle$ $s$--wave state by the counting rules.

%%%%

\acknowledgments
We thank Richard Lebed and Jose R. Pel\'aez for helpful discussions.
Work supported by Spanish grant MINECO:FPA2016-75654-C2-1-P, and by the 
US Department of Energy Contract No. DE-AC02-76SF00515.  SLAC-PUB-17340

%%%%%%%%%%%%%%%%%%%%%%%%%%%%%%%%%%%%%%%%%%%% 

\end{document}